\pdfoutput=1
\documentclass{sig-alternate-05-2015}

\usepackage{amsfonts,amssymb,amsmath,bm,graphicx,subfigure,makecell,hhline}
\usepackage{url,multirow,tabulary,color,float,pdfpages}
\newcommand{\nop}[1]{}

\newcommand{\delete}[1]{}
\newcommand{\add}[1]{{\bf \color{blue} [#1]}}

\begin{document}

\title{Ideology Detection for Twitter Users with Heterogeneous Types of Links}



%
%
%
%
%

\numberofauthors{2}
\author{
	\alignauthor Yupeng Gu \ \ \ Ting Chen \ \ \ Yizhou Sun \\
	\affaddr{University of California, Los Angeles, CA} \\
	\email{\{ypgu,tingchen,yzsun\}@cs.ucla.edu} 
	\alignauthor Bingyu Wang \\
	\affaddr{Northeastern University, Boston, MA} \\
	\email{rainicy@ccs.neu.edu} 
}

\maketitle

\begin{abstract}
The problem of ideology detection is to study the latent (political) placement for people, which is traditionally studied on politicians according to their voting behaviors. Recently, more and more studies begin to address the ideology detection problem for ordinary users based on their online behaviors that can be captured by social media, e.g., Twitter. As far as we are concerned, however, the vast majority of the existing methods on ideology detection on social media have oversimplified the problem as a binary classification problem (i.e., liberal vs. conservative). Moreover, though social links can play a critical role in deciding one's ideology, most of the existing work ignores the heterogeneous types of links in social media. In this paper we propose to detect \emph{numerical} ideology positions for Twitter users, according to their \emph{follow}, \emph{mention}, and \emph{retweet} links to a selected set of politicians. A unified probabilistic model is proposed that can (1) explain the reasons why links are built among people in terms of their ideology, (2) integrate heterogeneous types of links together in determining people's ideology, and (3) automatically learn the quality of each type of links in deciding one's ideology. Experiments have demonstrated the advantages of our model in terms of both ranking and political leaning classification accuracy. It is shown that (1) using multiple types of links is better than using any single type of links alone to determine one's ideology, and (2) our model is even more superior than baselines when dealing with people that are sparsely linked in one type of links. We also show that the detected ideology for Twitter users aligns with our intuition quite well.

\end{abstract}




\keywords{Ideology detection, heterogeneous network, social network analysis}

\section{Introduction}\label{sec:intro}

Ideology detection, i.e., ideal point estimation, dates back to early 1980s, where political scientists first cast light on ways of revealing politicians' political affiliation using their roll call voting data \cite{poole1985spatial}. Ideal points can be regarded as embeddings of political figures into a latent Euclidean space, and the positions are good representations of their political ideologies. Decades later, the estimation of ideal points has attracted interest from the computer science domain. Researchers begin to use more complicated, statistical models to conduct the estimation on roll call data \cite{gerrish2011predicting,gerrish2012they,gu2014topic}.

Recently, more and more studies pay attention to ideology detection for users on social media, which captures rich information for ordinary citizens in addition to political figures.  For example, based on text information, topic models and sentiment analysis approaches are utilized to classify users based on their opinions on different topics \cite{pla2014political, gottipati2013predicting}, and sometimes text features are directly fed into standard classification models such as support vector machines and boosted decision trees to classify users' political affiliations \cite{conover2011predicting, rao2010classifying, pennacchiotti2011machine}. Based on users' public profile information, such as marriage status, age and income, classifiers are trained to get users' ideology category \cite{al2012homophily}. Based on link information, various network-based methods are proposed to infer user's ideology \cite{golbeck2011computing, wong2013quantifying, wong2013media, zhou2011classifying, barbera2015birds}.

However, there are two major limitations of the existing literature. 
First, most of these approaches oversimplify the ideology detection problem as a binary classification problem, i.e., classify people as liberal or conservative, ignoring the fact that people's ideology lies in a very broad spectrum. For example, from our estimations in Fig. \ref{fig:intro-cs} we can see that Anderson Cooper (@andersoncooper) is almost in the middle and only a little bit to the left, while Rachel Maddow (@maddow) is much more liberal. 
Second, despite the successful utilization of link information in determining one's ideology, most of the works ignore the heterogeneous link types in social media, which leads to significant information loss. 
As shown in Fig. \ref{fig:intro-cs}, by looking only at the \emph{follow} links, there is no way to differentiate the ideology between Rachel Maddow (@maddow) and Megyn Kelly (@megynkelly); however from their \emph{retweet} links, it is quite clear the two have very different ideology. 
Take Glenn Greenwald (@ggreenwald) as another example: he has equal number of links to Barack Obama (@BarackObama) and Rand Paul (@RandPaul) in our network.
Only by knowing \textit{retweet} is a more important link type than \textit{follow}, we are able to predict Greenwald's ideology as liberal. It requires a model that can integrate various link types (i.e., relations) and assign them with different weights to provide a better overall ideology estimation. Admittedly, we may pre-define weights on different link types so as to obtain an integrated weighted graph, but it either requires strong domain knowledge on defining the link weights, or takes huge amount of time to do multiple rounds of training and select the set of weights with the best held-out performance, which makes it infeasible for real world networks. 


\begin{figure}[!htbp]
	\centering
	\includegraphics[width=1.0\linewidth]{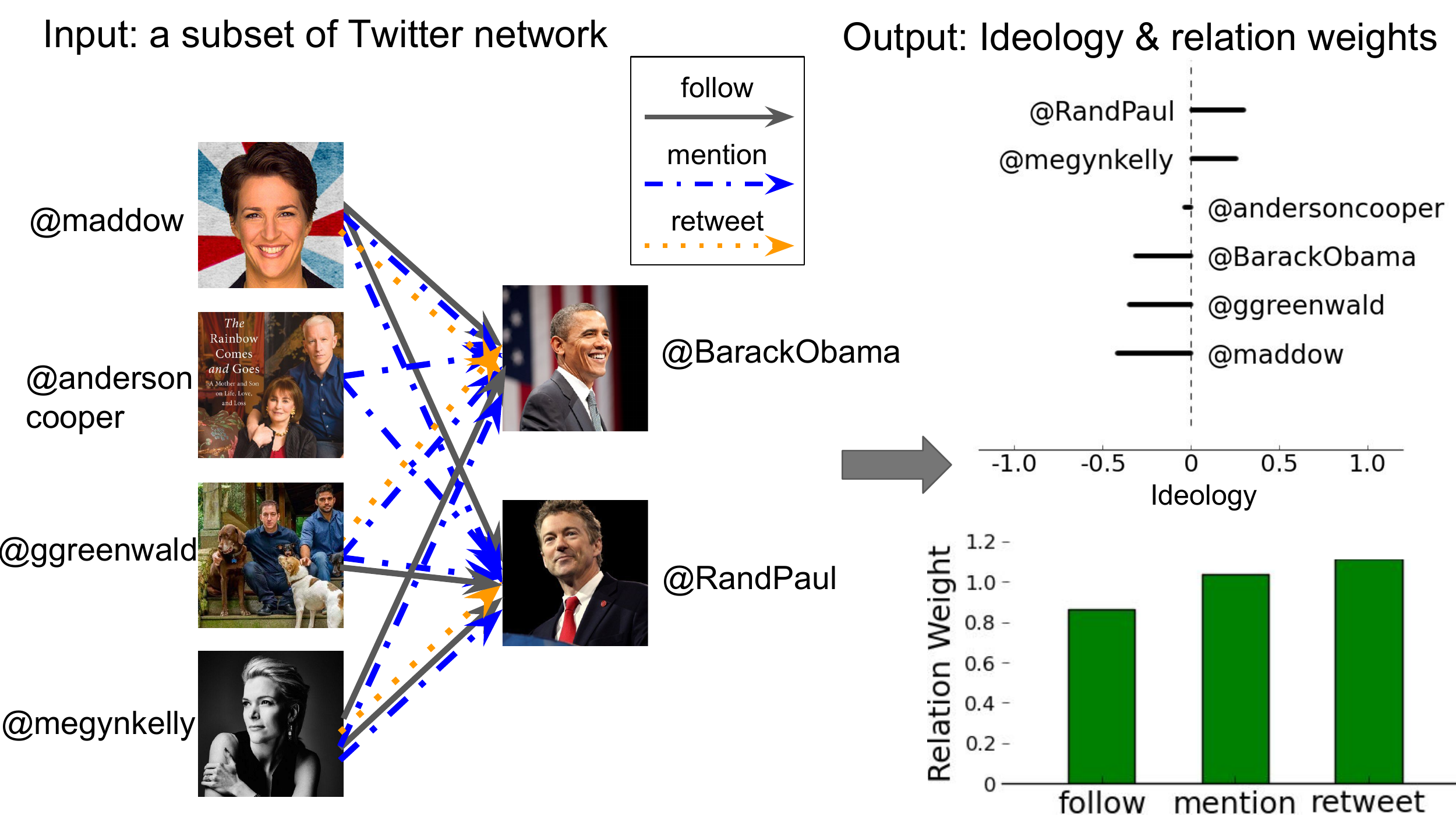}
	\caption{An illustration of ideology detection problem in Twitter network.}
	\label{fig:intro-cs}
\end{figure}

\nop{
\begin{figure}[h!tbp]
	\centering
	\includegraphics[width=1.0\linewidth]{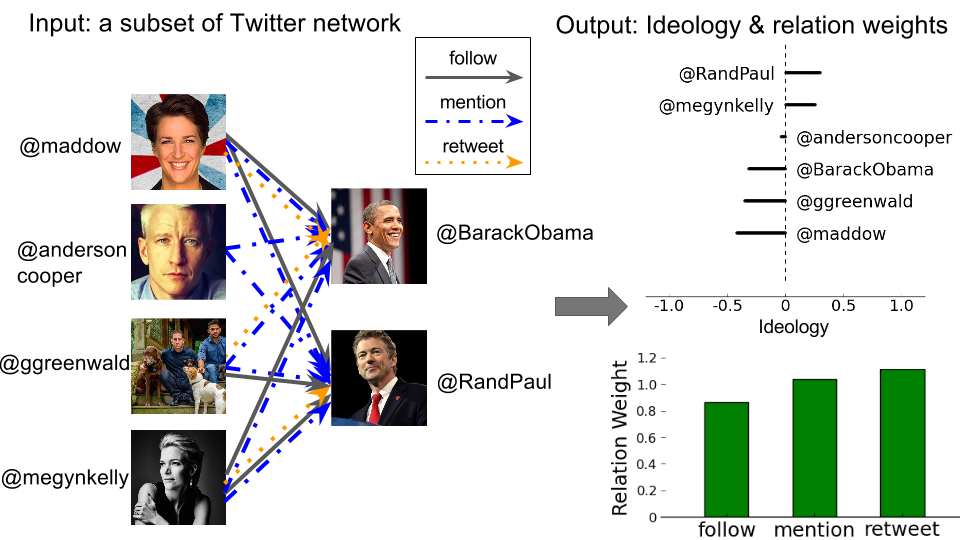}
	\caption{An illustration of ideology detection problem in Twitter network.}
	\label{fig:intro-cs}
\end{figure} 
}


In this paper, we propose a unified probabilistic model, \emph{ML-IPM} (\underline{M}ulti-\underline{L}ink \underline{I}deal \underline{P}oint Estimation \underline{M}odel), to detect \emph{numerical} ideology positions for a subset of Twitter users, according to their \emph{follow}, \emph{mention}, and \emph{retweet} links. Our approach has several advantages. First, it utilizes social choice theory to explain the reasons why links are built among people in terms of their ideology. Second, it is able to combine heterogeneous types of links in determining people's ideology, with different weights for each link type. Third, the strength of each link type can be automatically learned according to the observed network. Experiments demonstrate the advantages of our model in terms of both ranking and political leaning prediction accuracy. It is shown that (1) using multiple types of links is better than using any single type of links alone to determine one's ideology, and (2) our model is even better than baselines when dealing with people that are sparsely linked in one type of links. We also show that the detected ideology for Twitter users aligns with our intuition quite well. 
The main contributions of our work are summarized as below:
\begin{itemize}
	\itemsep0em
	\item We propose to detect \emph{numerical} ideology for Twitter users using link information only. In particular, heterogeneous types of links including \emph{follow}, \emph{mention}, and \emph{retweet} are considered. 
	\item We propose a unified probabilistic model, \emph{ML-IPM}, to solve the problem, which can (1) explain why links are built according to social choice theory and (2) consider different types of links by modeling their strength in determining one's ideology.
	\item A scalable learning algorithm is provided to learn both the ideology of users and the strength of each link type. Experimental results on a Twitter subnetwork demonstrate the advantages of our model over the state-of-the-art approaches. 
\end{itemize}

\section{Preliminaries and Problem Definition} \label{sec:prelim}

In this section, we introduce some preliminaries and define our problem.

\subsection{Ideal Point / Ideology}
Ideal point (also known as political ideology) is a political term which indicates the political leaning of politicians and provides quantitative measures of congressmen across time \cite{poole1985spatial,poole1991,Poole1997}. \nop{Literally, the ideology of a congressman can reflect his ideas on what he considers as the best form of capitalism and socialism. }It is a real-valued vector which refers to the position on the political spectrum, a real line or a high dimensional space that symbolizes political dimensions. In U.S. people usually refer ``left" as democracy or liberality, and ``right" as conservatism. Therefore, in one dimensional case, a person whose ideal point is negative (left of the origin) is treated as liberal. The more negative a person's ideal point is, the more liberal he/she is considered to be. Ideology is believed to be an important measure of political behaviors, and ideology estimation is essential in many political science applications. 

\delete{
\subsection{Twitter Network}
\delete{As one of the most influential online social networking outlets, Twitter has gained over hundreds of millions active registered users over the years\footnote{\url{https://about.twitter.com/company}}.} 
\add{On Twitter,} Users are allowed to \textit{follow} others, \textit{mention} others in their own tweets and \textit{retweet} others' tweets. \delete{Specifically, when a user $u_i$ follows $v_j$, we call $v_j$ the followee of $u_i$, and $u_i$ the follower of $v_j$. }In our framework, Twitter is regarded as a directed graph, where users are considered as nodes and ``\textit{follow}", ``\textit{mention}", ``\textit{retweet}" relations form the edges of the network. 

People may follow/mention/retweet others due to various reasons, in addition to political reasons. As we aim at detecting political ideology for users, we follow the following procedure to construct a Twitter subnetwork where the nodes and links are related to politics to some extent. First, we create a set of Twitter users that are politicians; then we select the candidate users based on the number of politicians they follow; and finally we construct a subnetwork between candidate users and politicians. More detailed description can be found in Section \ref{sec:exp:data}.
}

\subsection{Problem Definition}
Before the definition of our problem, we first look at some vivid cases in order to motivate our work. Suppose $P_1$ and $P_2$ are two political parties in the following examples. 

\textit{Case 1.} User U is a fan of politics and she is extremely interested in political news. Because politicians in $P_1$ tweet more frequently and U wants to keep track of the latest news and policies, she decides to follow most of the politicians of party $P_1$ but a small fraction of politicians in $P_2$. However, most of her retweets come from people in $P_2$, as she is more happy about their words and opinions. Given the \textit{follow} and \textit{retweet} links from U to these politicians, can we predict which party does she agree with more? More specifically, can we provide U's relative position among all the users that appeared in the network?

\textit{Case 2.} User V hates $P_1$, and he will argue with politicians of $P_1$ by \textit{mentioning} them on Twitter to show his displeasure. Sometimes he also mentions politicians in $P_2$ to support their policies. As expected, V follows most politicians in $P_2$, but few politicians in $P_1$. In this scenario, can we predict that the position of V lies close to the positions of people in $P_2$ on the political spectrum, based on the \textit{follow} and \textit{mention} links issued by him?

In this work, we focus on the problem of ideology detection for Twitter users. Formally, given a heterogeneous Twitter subnetwork with $R$ types of links ($R=3$ in our case), our goal is to (1) learn the multidimensional ideology for all users presented in the network and (2) determine the strength of each type of links in the process of ideology detection, which can best explain the observed network.
Although defined on Twitter network, our approach is very general to other social networks. 
\delete{Notations used in our paper are summarized in Table \ref{table:notations}. 
\begin{table}[htbp]
	\centering
	\begin{tabular}{|c|l|} \hline
	$N_1$ & Total number of link-senders \\ \hline
	$N_2$ & Total number of link-receivers \\ \hline
	$K$ & The dimension of ideology \\ \hline
	$R$ & Total number of types of links in the network \\ \hline
	$\bm{p}_i$ & $K$-dimensional ideology for user $u_i$ \\ \hline
	$\bm{P}$ & \makecell[l]{$N_1 \times K$ ideology matrix containing ideology \\ for all users with at least one outgoing link} \\ \hline
	$w_r$ & Weight of link type $r$ \\ \hline
	$S_+^{(r)}$ & The set of existing links of type $r$ \\ \hline
	$S_-^{(r)}$ & The sampled set of non-existing links of type $r$ \\ \hline
	${e_{ij}^{(r)}}^+$ & \makecell[l]{The number of links from $u_i$ to $v_j$ \\of link type $r$ in $S_+^{(r)}$} \\ \hline
	${e_{ij}^{(r)}}^-$ & \makecell[l]{The number of non-existing links from $u_i$ to $v_j$ \\of link type $r$ in $S_-^{(r)}$} \\ \hline
	\end{tabular}
	\caption{\delete{Table of Notations}} 
	\label{table:notations}
\end{table}
}

\section{Approach} 
\delete{
There are three challenges for the ideology detection problem: 
\begin{itemize}
	\item How can we estimate users' political ideology on a social network using purely the links of the network? How does ideology help to explain the reason behind the generation of social links?
	\item For a heterogeneous network with multiple types of links, how to leverage information from different link types to achieve a better estimation of users' ideology?
	\item How we can automatically learn the strength associated with each type of links in determining one's ideology?	
\end{itemize}
We will discuss our approach that tackles these challenges thoroughly in this section. 
}


In this section, we introduce our solution to the proposed problem. We start from ideology model under a single link type, then introduce how to extend the model when multiple types of links exist, and finally introduce the learning algorithm.

\subsection{Ideology Estimation Model via Single Link Type} \label{sec:single}
\delete{Due to the non-reciprocal nature of Twitter links, Twitter is regarded as a directed network and every user can be thought of as both a link sender (e.g. follower) and a link receiver (e.g. followee).} As in traditional ideal point models, each user has a position in a $K$-dimensional space, which  represents his/her ideology. For a politics-related network, ideology can help explain the reason for link generation, which is a reflection of people's online behaviors. Take \textit{follow} link as an example: the proximity of two users' positions in the latent ideology space indicates a high probability that they have many politician friends (followees) in common, and vice versa. \delete{In other words, ideology should well reflect a user's taste in choosing friends. }Because Twitter users are free to follow anyone they like, we assume that each possible \textit{follow} link from $u_i$ to $v_j$ is the outcome of $u_i$'s choice of following $v_j$ or not. Each choice corresponds to a utility function which measures a user's preference over the two choices. Rational users are expected to \nop{interact with people that maximize their experience on social networks; in other words, they will }make the decision that maximizes their utility. Intuitively, the utility function should consider the distance between the positions of two users in the latent ideology space, since people are more likely to follow others whose political opinions are close to themselves. \nop{Following the similar idea of lawmakers' activity on bills \cite{poole1985spatial}, }We discuss why a link is formed from both the link-sender and link-receiver's points of view:

\textbf{The role of link-sender.} From a link-sender's (say user $u_i$) point of view, one has to make a choice on whether to follow user $v_j$ or not, which depends on the two utility functions defined on the two options. She will choose the option that maximizes her utility function, which is related to her ideology position $\bm{p}_i \in \mathbb{R}^{K}$. \delete{Note that in our Twitter subnetwork, link-senders will be the candidate users.}

\textbf{The role of link-receiver.} From a link-receiver's (say user $v_j$) point of view,  each of them is associated with two placements, corresponding to the two options of whether she will receive a link from a link sender or not. We use $\bm{\psi}_j \in \mathbb{R}^{K}$ and $\bm{\zeta}_j \in \mathbb{R}^{K}$ to denote the positions of the two placements, which can be observed by other users. \delete{Note that in our Twitter subnetwork, link-receivers will be the politicians.}

We now define the utility functions for a user $u_i$'s choice of (1) ``following'' a target user $v_j$ ($U_i(\bm{\psi}_j)$) and (2) not ``following'' a target user $j$ ($U_i(\bm{\zeta}_j)$), by extending the model in \cite{clinton2004statistical} to the scenario of social networks:
\begin{equation}
	\small
	\begin{aligned}
		U_i(\bm{\psi}_j) = u_i(\bm{\psi}_j) + \eta_{ij}, & \text{ if } u_i \text{ follows } v_j \\
		U_i(\bm{\zeta}_j) = u_i(\bm{\zeta}_j) + \nu_{ij}, & \text{ if } u_i \text{ does not follow } v_j
	\end{aligned}
\end{equation}
where $u_i(\bm{\psi}_j) = -||\bm{p}_i - \bm{\psi}_j||^2$, $u_i(\bm{\zeta}_j) = -||\bm{p}_i - \bm{\zeta}_j||^2$, and $\eta_{ij}$, $\nu_{ij}$ are random noises drawn from the standard Gumbel distribution, which is commonly used in discrete choice theory \cite{ben1985discrete}. In other words, the two utility functions are defined as the negative of squared distance function between user $u_i$'s ideology position and the positions of the two possible placements of user $v_j$, plus a random noise. 
 
According to the utility functions, user $u_i$ will follow $v_j$ if $U_i(\bm{\psi}_j) > U_i(\bm{\zeta}_j)$. Using the same manipulation in \cite{clinton2004statistical}, the probability that user $u_i$ decides to follow user $v_j$ is given by
\begin{equation} \label{equ:basicidp}
	\small
	p(u_i \to v_j) =  P(U_i(\bm{\psi}_j) > U_i(\bm{\zeta}_j)) = \sigma(\bm{p}_i \cdot \bm{q}_j + b_j) 
\end{equation}
where $\bm{q}_j = 2(\bm{\psi}_j - \bm{\zeta}_j)$, $b_j = ||\bm{\zeta}_j||^2 - ||\bm{\psi}_j||^2$ and $\sigma(x) = \frac{1}{1+e^{-x}}$ is the sigmoid function. $\bm{q}_j$ can be interpreted as the image of user $v_j$ viewed by others, and $b_j$ can be regarded as a bias term for $v_j$. Usually we can only estimate $\bm{q}_j$ and $b_j$ instead of $\bm{\psi}_j$ and $\bm{\zeta}_j$, and it is adequate since we are interested in the ideology $\bm{p}_i$ and the probability of a link can be calculated based on $\bm{q}_j$ and $b_j$. 

After the probability of a directed link between two users is determined, we are able to write down the log-likelihood of observing a network $G$ as 
\begin{equation} \label{equ:verbose_logl}
	\small
	\begin{aligned}
		l(G) &= \log \Big( \prod_{(i,j)}
		 \sigma_{ij}^{I_{[i \to j]}} (1-\sigma_{ij})^{1-I_{[i \to j]}} \Big) \\
		&= \sum_{\substack{(i,j) : i \to j}} \log \sigma_{ij}
		 + \sum_{\substack{(i,j) : i \nrightarrow j}} \log (1-\sigma_{ij})
	\end{aligned}
\end{equation}
where $I_{[\cdot]}$ is the indicator function, $i \to j$ means there is a link from $u_i$ to $v_j$, and $i \nrightarrow j$ means there are no links from $u_i$ to $v_j$. Here $\sigma_{ij}$ is short for $\sigma(\bm{p}_i \cdot \bm{q}_j + b_j)$ in Eq. \ref{equ:basicidp}.

Obviously not every pair of people has a link in between. In fact, links in social networks are rather sparse. Therefore we will not go through every possible pair of links in the likelihood above. Instead, we select all the existing links and sample the same number of non-existing links. We denote the set of existing links as $S_+$ and the sampled set of non-existing links as $S_-$. \delete{Hence Eq. \ref{equ:verbose_logl} becomes 
\begin{equation}
	\small
	l(G) = \sum_{(i,j) \in S_+} \log \sigma_{ij} 
	 + \sum_{(i,j) \in S_-} \log (1-\sigma_{ij})
\end{equation}
}
In addition, in the case where multiple links may exist between users (e.g., a user can \textit{mention} another user several times), our model can be generalized by simply treating multiple links as multiple occurrences of a single edge. Suppose $e_{ij}^+$ is the number of links from $u_i$ to $v_j$ and $e_{ij}^-$ is the number of sampled non-existing links from $u_i$ to $v_j$ (typically $e_{ij}^-=1$), then the log-likelihood above becomes
\begin{equation}
	\small
	l(G) = \sum_{(i,j) \in S_+} e_{ij}^+ \cdot \log \sigma_{ij} 
	 + \sum_{(i,j) \in S_-} e_{ij}^- \cdot \log (1-\sigma_{ij})
\end{equation}

\subsection{Ideology Estimation Model via Multiple Link Types}

We now address the challenge of utilizing multiple types of links for ideology detection. In a heterogeneous network, nodes can be connected via different types of relations\nop{, and each type of relation can be considered as a distinct way of interaction between users}. On Twitter, people can \textit{follow}, \textit{mention} or \textit{retweet} others. It naturally forms three different types of links, and different link types certainly have different interpretations. 

In order to utilize the knowledge in the heterogeneous network across all kinds of relations, our ultimate model is a weighted combination of models using all types of links. According to our previous assumption, a link-sender $u_i$ will have the same intrinsic ideal points $\bm{p}_i$ across all link types. However we posit that the roles of link-receivers are different for different relations, as the images of users will change when observed by different types of behaviors. For example, $u_i$ can easily decide to \textit{follow} $v_j$ but hesitates to \textit{retweet} from $v_j$. Therefore we use the same $\bm{p}_i$ for user $u_i$ for all types of links, and modify $\bm{\psi}_j$, $\bm{\zeta}_j$ to $\bm{\psi}_j^{(r)}$, $\bm{\zeta}_j^{(r)}$ which are specific to the relation $r$. As a result, $\bm{q}_j$ and $b_j$ will be changed to $\bm{q}_j^{(r)}$ and $b_j^{(r)}$ as well. In consideration of the heterogeneity in different types of links, we also add a relation weight $w_r$ which represents the relative importance of the links in the corresponding relation type $r$. Besides, we will use the average log likelihood of each link in each type of relation in order to balance the scale of different link types.

Denoting $\bm{P} = \{\bm{p}_i\}, \bm{Q} = \{\bm{q}_j^{(r)}\}$ and $\bm{B} = \{b_j^{(r)}\}$, we define our objective as
\begin{equation} \label{equ:all_logl}
	\scriptsize
	\begin{aligned}
	& ~~~ l( \bm{G} | \bm{P}, \bm{Q}, \bm{B} ) \\
	=& \sum_{r=1}^{R} { w_r \cdot \frac{ \sum\limits_{(i,j) \in S_+^{(r)}} e_{r,ij}^+ \log{\sigma_{r,ij}}
	 + \sum\limits_{(i,j) \in S_-^{(r)}} e_{r,ij}^- \log{( 1-\sigma_{r,ij} )} } 
	 { \sum\limits_{(i,j) \in S_+^{(r)}} e_{r,ij}^+  
	  + \sum\limits_{(i,j) \in S_-^{(r)}} e_{r,ij}^- }
	}
	\end{aligned}
\end{equation}
where $e_{r,ij}^+$ is the number of existing links from $u_i$ to $v_j$ via relation $r$, $e_{r,ij}^-$ is the number of sampled non-existing links from $u_i$ to $v_j$ via relation $r$ and $\sigma_{r,ij} = \sigma(\bm{p}_i \cdot \bm{q}_j^{(r)} + b_j^{(r)})$ for short. We will use these notations in the remaining of this section.
The constraint we put on $w_r$ is $w_r > 0, ~r=1,\cdots,R$ and $\prod_{r=1}^{R}{w_r} = 1$, which forces the geometric mean\footnote{The geometric mean of several positive variables $w_1,\cdots,w_R$ is given by $(\prod_{r=1}^{R}{w_r})^{1/R}$.} of $w_1, \cdots, w_R$ to be 1. In the remaining of the paper we will use the $R$-dimensional vector $\bm{w}=(w_1,\cdots,w_R)$ to denote the vector of network weights. For succinctness, we will use $N_r$ to denote $\sum\limits_{(i,j) \in S_+^{(r)}} e_{r,ij}^+ + \sum\limits_{(i,j) \in S_-^{(r)}} e_{r,ij}^-$, i.e., the total number of links in relation $r$.

\textbf{Identification.} Because the probability in Eq. \ref{equ:basicidp} involves the product of two sets of parameters, there would be an identifiability issue where the objective will be exactly the same if we multiply every $\bm{p}_i$ by a non-zero constant $c$ and divide every $\bm{q}^{(r)}_j$ by the same constant. In other words we need to control the scale of model parameters $\bm{P}$ and $\bm{Q}$. \delete{In order to avoid such a problem, some researchers choose specific users as fixed anchors or apply a non-zero prior over them to encourage the model to prefer one of the parameters \cite{jackman2001multidimensional, clinton2004statistical}. Others prefer to add a Gaussian prior to all users or equivalently, add an $l_2$ regularization term over those parameters.} In our work we add an $l_2$ regularization term on $\bm{P}$ and $\bm{Q}$. We also add the same regularization on $\bm{B}$ to avoid overfitting. Therefore Eq. (\ref{equ:all_logl}) becomes 
\begin{equation} \label{equ:all_logl_reg}
	\scriptsize
	\begin{aligned}
	& ~~~ l( \bm{G} | \bm{P}, \bm{Q}, \bm{B} ) \\
	=& \sum_{r=1}^{R} { w_r \cdot \frac{ \sum\limits_{(i,j) \in S_+^{(r)}} e_{r,ij}^+ \log{\sigma_{r,ij}}
	 + \sum\limits_{(i,j) \in S_-^{(r)}} e_{r,ij}^- \log{( 1-\sigma_{r,ij} )} }
	 { N_r }
	} \\
	&- \frac{\mu}{2} \big( ||\bm{P}||_F^2 + \sum_{r=1}^{R} ||\bm{Q}^{(r)}||_F^2 + \sum_{r=1}^{R} ||\bm{b}^{(r)}||_2^2 \big)
	\end{aligned}
\end{equation}
Here $||\cdot||_F$ denotes the Frobenius norm of a matrix: $||A||_F = \Big( \sum_{m,n} A_{mn}^2 \Big)^{1/2}$ and $\mu>0$ is a parameter that controls the effect of regularization terms.

\subsection{Optimization}
We propose a two-step algorithm to optimize the model parameters. The first step is updating $\bm{w}$ fixing all other parameters. The second step is updating $\bm{P},\bm{Q},\bm{B}$ given $\bm{w}$. The two steps will run iteratively until convergence. 

\textbf{Update $\bm{w}$.} Fortunately we are able to find a closed form solution for network weights $\{w_r\}_{r=1}^{R}$. Maximizing the objective function (Eq. \ref{equ:all_logl_reg}) w.r.t $w_r$ is equivalent to minimizing 
\begin{equation} \label{equ:equi}
	\small
	\begin{aligned}
		J(\bm{w}) = \sum_{r=1}^{R} w_r \cdot L_r 
		~~~~\text{s.t. } \sum_{r=1}^{R} \log w_r = 0
	\end{aligned}
\end{equation}
where 
\begin{equation}
	\small
	\begin{aligned}
	L_r = - \frac{ \sum\limits_{(i,j) \in S_+^{(r)}} e_{r,ij}^+ \log{\sigma_{r,ij}}
	 + \sum\limits_{(i,j) \in S_-^{(r)}} e_{r,ij}^- \log{( 1-\sigma_{r,ij} )} } 
	 { N_r } > 0
	\end{aligned}
\end{equation}
is a constant with respect to $w_r$, which can be considered as the loss function associated with relation $r$.\delete{ The Lagrangian of Eq. \ref{equ:equi} is defined by 
\begin{equation}
	\small
	\mathcal{L}(\bm{w}, \lambda) = \sum_{r=1}^{R} w_r \cdot L_r + \lambda \cdot (-\sum_{r=1}^{R} \log w_r)
\end{equation}
}
We can find the optimal value of $w_r$ by setting the derivative of \delete{above }Lagrangian w.r.t. $w_r$ to zero, which leads to 
\begin{equation} \label{equ:lambda}
	\scriptsize
	\lambda = \Big( \prod_{r=1}^{R}{L_r} \Big)^ {1/R}
\end{equation}
and 
\begin{equation}
	\small
	w_r = \frac{(\prod_{r=1}^{R}{L_r})^{1/R}}{L_r}, ~~\forall r=1,\cdots,R
\end{equation}

The value of $w_r$ in each round can be interpreted as the geometric mean of the loss for all link types divided by the loss for link type $r$. Intuitively, in order to minimize the objective function, we should assign small weights to relations with big loss ($L_r$), namely smaller average log likelihood. In other words, we will assign a big weight to a relation that is consistent with the current model (i.e., with a big average log likelihood function), which agrees with our intuition. 

\textbf{Update $\bm{P},\bm{Q},\bm{B}$.} Given $\bm{w}$, all other parameters are updated using gradient ascent algorithm. \delete{The gradient with respect to all the parameters can be calculated as 
\begin{equation}
	\small
	\begin{aligned}
		\frac{\partial J}{\partial p_{ik}} & = 
		\sum_{r} \Big( \frac{w_r}{N_r} \cdot 
			\sum_{j: (i,j) \in S_+^{(r)}} e_{r,ij}^+ ~ q_{jk}^{(r)} ~ (1-\sigma_{r,ij}) \Big) \\
		& - \sum_{r} \Big( \frac{w_r}{N_r} \cdot 
			\sum_{j: (i,j) \in S_-^{(r)}} e_{r,ij}^- ~ q_{jk}^{(r)} ~ \sigma_{r,ij} \Big)
		- \mu \cdot p_{ik} \\
		\frac{\partial J}{\partial q_{jk}^{(r)}} & = 
		\frac{w_r}{N_r} \cdot \Big( 
			\sum_{i: (i,j) \in S_+^{(r)}} e_{r,ij}^+ ~ p_{ik} ~ (1-\sigma_{r,ij}) 
			\Big) \\
		& - \frac{w_r}{N_r} \cdot \Big( 
			\sum_{i: (i,j) \in S_-^{(r)}} e_{r,ij}^- ~ p_{ik} ~ \sigma_{r,ij} 
			\Big)
		- \mu \cdot q_{jk}^{(r)} \\
		\frac{\partial J}{\partial b_{j}^{(r)}} & = 
		\frac{w_r}{N_r} \cdot \Big( \sum_{i: (i,j) \in S_+^{(r)}} e_{r,ij}^+  ~ (1-\sigma_{r,ij}) \Big) \\
		& - \frac{w_r}{N_r} \cdot \Big( \sum_{i: (i,j) \in S_-^{(r)}} e_{r,ij}^- ~ \sigma_{r,ij} \Big)
		- \mu \cdot b_{j}^{(r)}
	\end{aligned}
\end{equation}
}


\textbf{Time complexity.} In each iteration the time complexity of updating $\bm{w}$ is \begingroup\makeatletter\def\f@size{7}\check@mathfonts $O(K \cdot \sum_{r=1}^{R} {N_r})$\endgroup, and updating $\bm{P},\bm{Q},\bm{B}$ requires \begingroup\makeatletter\def\f@size{7}\check@mathfonts$O(K \cdot \sum_{r=1}^{R} {N_r})$\endgroup operations as well, where $K$ is the dimension of the ideology vector. Therefore the time complexity of our model is linear to the number of edges in the network in particular. Hence, our proposed algorithm is efficient and can be scaled to large networks.

\nop{We adapt backtracking line search strategy \cite{armijo1966minimization} 
to make our updating process more efficient. }

\section{Experiments}
In this section we will show the experimental results based on a subset of Twitter network. We will first introduce how to construct the subnetwork. Then we will compare our method with several baselines, in terms of both ranking accuracy and political leaning classification. Finally, we conduct extensive case studies to show the estimated ideology is meaningful and agrees with our intuition very well. 

\subsection{Data Preparation}\label{sec:exp:data}
In order to prepare our dataset, we start from politicians and track the users that are connected to them. We first collect the list of all the members of the $113^{th}$ U.S. congress (2013-2015). We manually search their names on Twitter and match a name to an account if the account is verified and we have enough confidence to show he/she is the corresponding congressman/congresswoman. This step will provide us a list of 487 politicians. \delete{Note that in our training process we do assume all the labels of politicians are unknown.}

We then use Twitter's streaming API\delete{\footnote{https://dev.twitter.com/streaming/overview}} and REST API\delete{\footnote{https://dev.twitter.com/rest/public}} to collect their followees and followers. In consideration of efficiency, we collect at most 5,000 followers and followees for every congressman. On one hand, in order to select politics-related users, we set a threshold $t=20$ where we only keep users who follow at least $t$ congressmen or are followed by at least $t$ congressmen. These accounts are likely to be enthusiastic about political issues. On the other hand, we also include around 10,000 random users who follow 3$\sim$5 politicians to include more peripheral (less politics-related) Twitter users. \nop{Two groups of users are called core users and peripheral users, respectively. These two procedures provide us with a total of 46,477 users in the dataset. }Our approach will be evaluated on this Twitter subnetwork with these users as vertices. 

There are various reasons apart from ideology why people interact with each other on Twitter, such as they know each other in real life. Therefore, preparing the set of link-receivers is crucial in that the decision to follow them or not should reveal other users' political interests. In order to alleviate the effect of random factors in our model, we restrict the set of link-receivers to be U.S. congress members; in other words, the social links between two ordinary citizens will not be taken into consideration. This strategy allows us to minimize other social factors behind link generation in our model, and it will make our model more accurate. 

Here is how we construct the social network for different relations: for \textit{follow} link type, we will add an edge from $u_i$ to $v_j$ to the network if a user $u_i$ follows a politician $v_j$. For \textit{mention} and \textit{retweet} link types, we collect part\footnote{At most 3,200 tweets for each user are available due to API limits.} of recent tweets for all the users. \delete{The crawling process was done between April 2014 and January 2016. If $u_i$ posts a tweet which mentions a politician $v_j$, we add an edge from $u_i$ to $v_j$ to the \textit{mention} network. Similarly, if $u_i$ retweets a tweet from a politician $v_j$, we will add an edge from $u_i$ to $v_j$ to the \textit{retweet} network.}If $u_i$ posts a tweet which mentions/retweets from a politician $v_j$, we add an edge from $u_i$ to $v_j$ to the \textit{mention}/\textit{retweet} network. If a user does not mention (retweet from) any politician, he/she will be an isolated node and we will simply remove the user from the corresponding network. Note that in contrast to the follow links, a user can mention or retweet another user for multiple times. In this case we simply assume multiple edges exist between them, as discussed in Section \ref{sec:single}. More details of our dataset can be found in Table \ref{table:stat}. 
\begin{table}[!h]
	\small
	\centering
	\begin{tabulary}{0.5\textwidth}{|c||c|c|c|} \hline
	Relation & \textit{follow} & \textit{mention} & \textit{retweet}  \\ \hline
	Number of links & 1,764,956 & 2,395,813 & 718,124 \\ \hline
	Total number of users & 46,477 & 34,775 & 30,990 \\ \hline
	\end{tabulary}
	\caption{Statistics for Twitter Dataset}
	\label{table:stat}
\end{table}

\subsection{Performance Evaluation}
\subsubsection{Baseline Methods}
We compare our Multiple Link Types Ideal Point Estimation Model (ML-IPM) with the following baseline methods: 
\begin{itemize}
	\item The simplest baseline where the ideology of a user is the average score of her outgoing neighbors. Each Republican is assigned an ideology score of 1, and each Democrat is assigned a score of -1. We denote this baseline method as AVER.
	\item The Bayesian Ideal Point Estimation Model (B-IPM) \cite{barbera2015birds} to be introduced in Section \ref{sec:related}. Although the author does not mention their generalization to relations other than \textit{follow}, we adopt the model for other types of links for comparison. 
	\item Our Single Link Type Ideal Point Estimation Model (SL-IPM) where only one type of link is present in the social network, as introduced in Section \ref{sec:single}. 
	\item A special case of our model ML-IPM where the weights for different types of links are fixed (ML-IPM-fixed). In this case the weights for different link types are uniformly distributed, namely $w_1=w_2=\cdots=w_R=1$. 
\end{itemize}

\delete{By comparing our model with the first two baselines, we are able to show the strength of our approach over the state-of-the-art literature. Improvement over the third baseline demonstrates the significance in utilizing information from various sources in the social network, and the comparison to the last baseline illustrates our strength in automatically learning the importance score of multiple relations. }The research problem in \cite{wong2013quantifying} is also similar to our task (with single link type), however their approach is not scalable to the size of our dataset (the time complexity is $O(n^2)$ for each iteration, where $n$ is the number of users). Therefore we will not consider their method for comparison. 

\subsubsection{Evaluation Measures}
In our experiments, we will evaluate the ranking and classification accuracy to demonstrate the effectiveness of our ideal point estimation model.  

\textbf{Ranking.} In order to evaluate the effect of continuous ideology, we design the ranking evaluation based on 100 manually labeled users, with integer labels from 1 (most liberal) to 5 (most conservative). Here we evaluate the pairwise accuracy between Twitter users, where a pair of users is considered correct if the order of their 1-dimensional ideologies aligns with the order of manual labels (for example, a pair of users $(u_a,u_b)$ where $u_a$ is labeled as ``1'' and has an ideology of $-0.9$, and $u_b$ is labeled as ``4'' and has an ideology of $0.2$). We use five different sets of random initialization for the model parameters, and report the mean and standard deviation on a total of 3,857 pairs of users.

\textbf{Classification.} In the classification task, we will classify users as liberal or conservative based on the ideology we have inferred from the dataset. To obtain the ground truth of some users in our dataset, we crawl congress people's party affiliation from the Internet (which is open to public). In addition, we collect the political bias for 100 popular newspaper accounts\footnote{Source: \url{http://www.mondotimes.com/newspapers/usa/usatop100.html}}, and we also take advantage of the labeled users in our previous task. These multi-dimensional ideal points are used to train a logistic regression classifier. The classification performance is measured by the Area Under ROC Curve (AUC), and is averaged over 10 different runs by sampling different training data. For all the experiments, we use five different sets of random initialization for the model parameters. 
\begin{table}[!h]
	\centering
	\begin{tabular}{|c|c|c|} \hline
		Method & Ranking Accuracy & Classification AUC \\ \hline \hline
		AVER (\textit{follow}) & 0.427 & 0.523 \\ \hline
		AVER (\textit{mention}) & 0.446 & 0.558 \\ \hline
		AVER (\textit{retweet}) & 0.474 & 0.587 \\ \hline \hline
		B-IPM (\textit{follow}) & $0.443 \pm 0.102$ & $0.868 \pm 0.021$ \\ \hline
		B-IPM (\textit{mention}) & $0.433 \pm 0.183$ & $0.558 \pm 0.064$ \\ \hline
		B-IPM (\textit{retweet}) & $0.501 \pm 0.127$ & $0.561 \pm 0.066$ \\ \hline \hline
		SL-IPM (\textit{follow}) & $0.626 \pm 0.011$ & $0.953 \pm 0.015$ \\ \hline
		SL-IPM (\textit{mention}) & $0.623 \pm 0.027$ & $0.951 \pm 0.018$ \\ \hline
		SL-IPM (\textit{retweet}) & $0.637 \pm 0.005$ & $0.958 \pm 0.005$ \\ \hline \hline
		\makecell{ML-IPM-fixed} & $0.655 \pm 0.008 $ & $0.930 \pm 0.035$ \\ \hline
		\makecell{ML-IPM} & $\bm{0.663 \pm 0.007}$ & $\bm{0.986 \pm 0.013}$ \\ \hline
	\end{tabular}
	\caption{Experimental Results} 
	\label{table:ranking_classification}
\end{table}

\nop{
\begin{table}[!h]
	\centering
	\begin{tabular}{|c|c|} \hline
		Method & Classification AUC \\ \hline \hline
		AVER (\textit{follow}) & 0.5227 \\ \hline
		AVER (\textit{mention}) & 0.5582 \\ \hline
		AVER (\textit{retweet}) & 0.5871 \\ \hline \hline
		B-IPM (\textit{follow}) & $0.8676 \pm 0.0650$ \\ \hline
		B-IPM (\textit{mention}) & $0.5583 \pm 0.2005$ \\ \hline
		B-IPM (\textit{retweet}) & $0.5608 \pm 0.2100$ \\ \hline \hline
		SL-IPM (\textit{follow}) & $0.9532 \pm 0.0477$ \\ \hline
		SL-IPM (\textit{mention}) & $0.9505 \pm 0.0577$ \\ \hline
		SL-IPM (\textit{retweet}) & $0.9577 \pm 0.0173$ \\ \hline \hline
		\makecell{ML-IPM-fixed} & $0.9304 \pm 0.1114$ \\ \hline
		\makecell{ML-IPM} & $\bm{0.9866 \pm 0.0453}$ \\ \hline
	\end{tabular}
	\caption{Evaluation on classification.}
	\label{table:classification}
\end{table}
}

\textbf{Discussion.} 
We report the ranking accuracy and classification AUC in Table \ref{table:ranking_classification}, where the relation in the bracket represents the type of link used in the corresponding method. From the above table we can see our advantage over the baseline methods in terms of user ranking and classification accuracy. The comparison on a single link type between our approach (rows 7-9) and existing work (rows 1-3,4-6) illustrates our considerable improvement for both tasks. Nevertheless, information from a single network may not be adequate in determining one's ideology: people may \textit{follow} well-known political figures for reasons other than political proximity, and it is not rare that a user keeps criticizing others by \textit{mentioning} them, or \textit{retweets} ironically. Therefore, integrating heterogeneous types of links becomes necessary, and the results are always better if we automatically update the weight for each link type (rows 10-11). In addition, our method is not sensitive to initialization as the standard deviation is small (row 11). In sum, our learned ideology has a higher quality when we combine heterogeneous types of links together, with learned weights of all link types.

\subsubsection{Cold-start Problem}
Cold start problem is known as the issue in a system where the inference for users without sufficient information will suffer. Our unified approach is also able to tackle the cold-start problem by integrating knowledge from other types of links: users that are sparsely linked by one type of link may be active in other networks. Thus we are able to have a significantly better understanding of the users by utilizing all types of links. 

\begin{figure*}[!t]
	\begin{center}
	\begin{minipage}{0.4\textwidth}
		\centering
		\includegraphics[width=\linewidth]{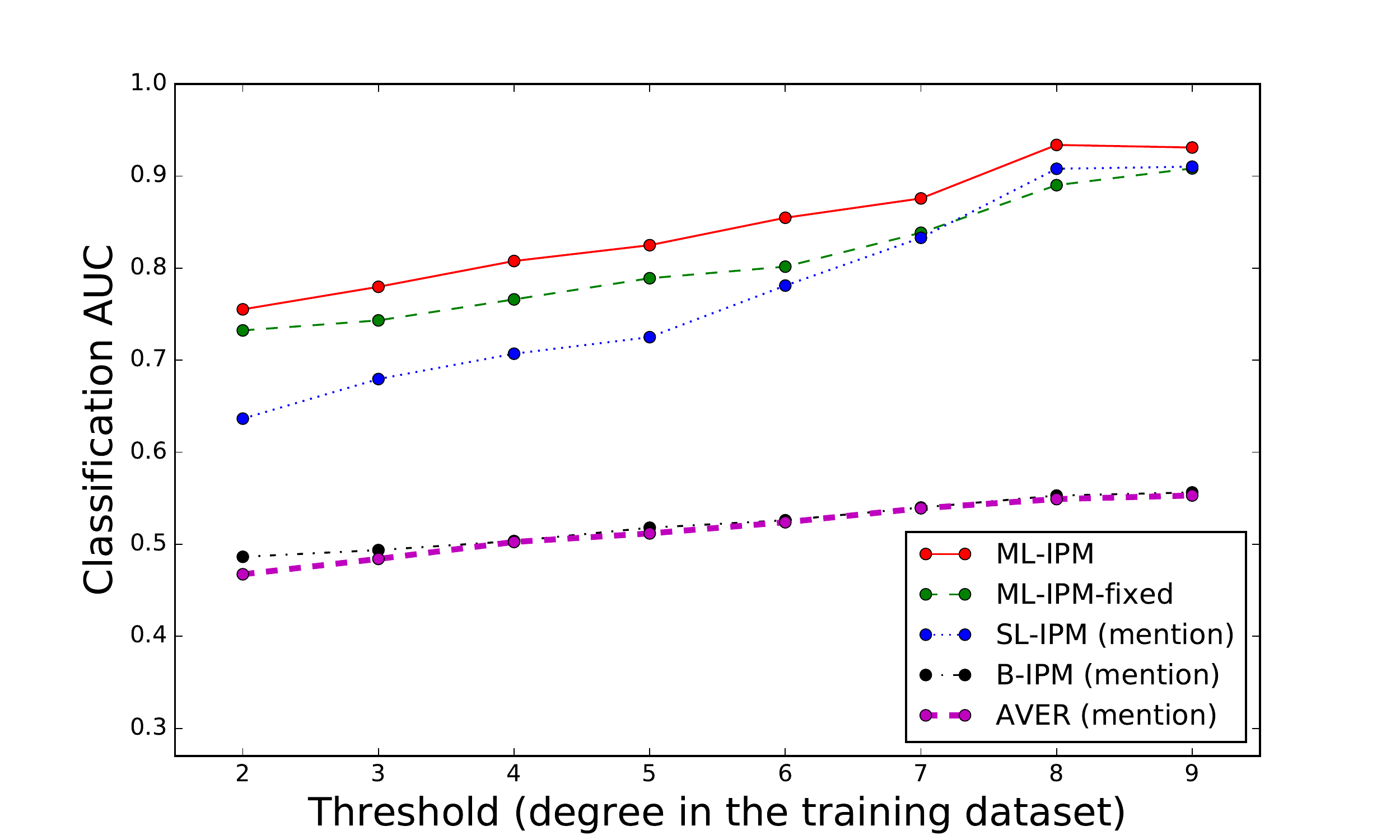} \\
		\textit{mention} network
	\end{minipage}%
	\begin{minipage}{0.4\textwidth}
		\centering
		\includegraphics[width=\linewidth]{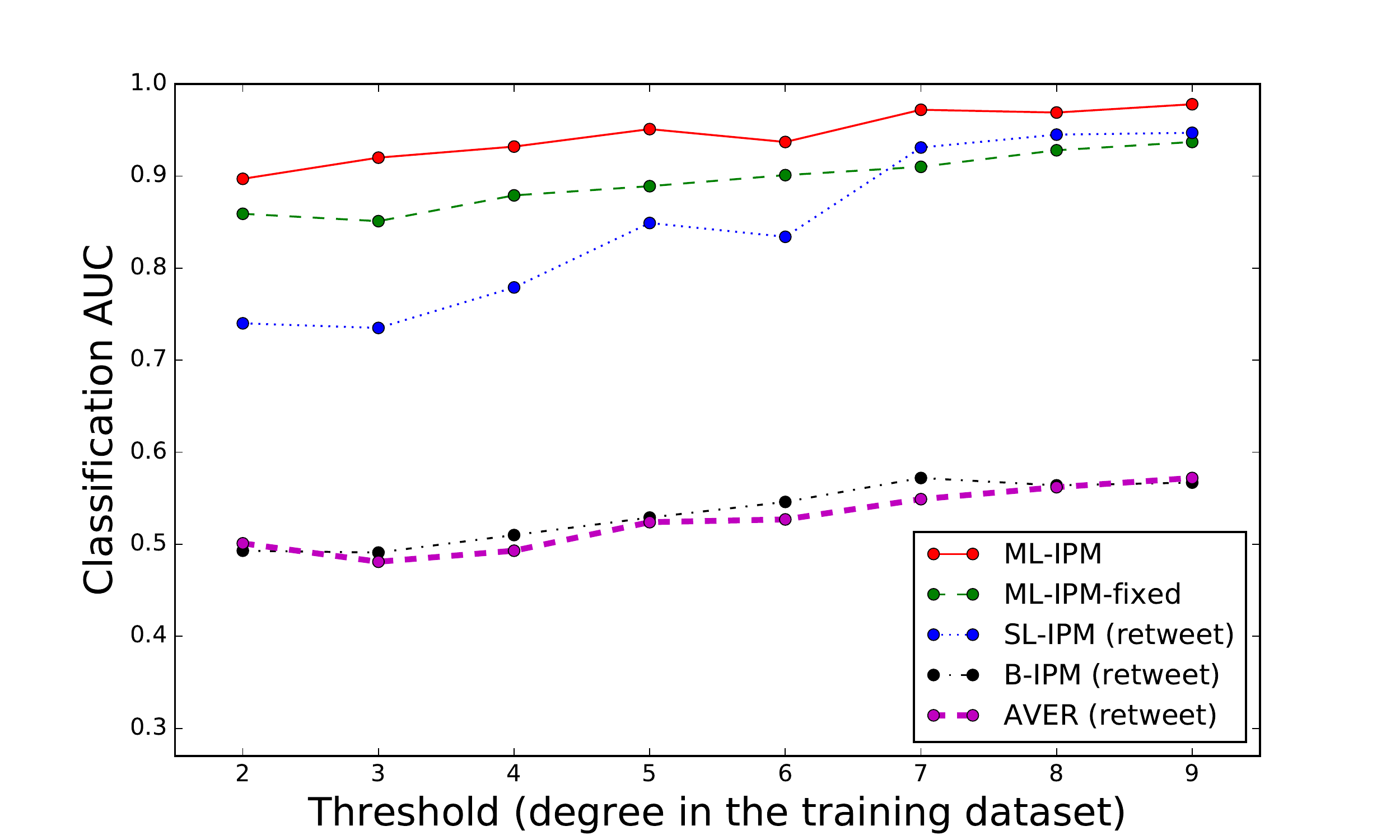}  \\
		\textit{retweet} network
	\end{minipage}%
	\end{center}
	\caption{Classification AUC for sparsely linked users in different types of networks.}
	\label{fig:cold_start_classification}
\end{figure*}

To prepare the candidate users, we select all users whose number of outgoing links is less than or equal to a given threshold in the training dataset, and we use classification AUC (as in previous subsection) and link prediction as evaluation measures. In the link prediction task, we observe some social links in the network and predict a set of unobserved links of the same type. In the experiments, we also use five different initializations and 90\% of the links (both existing and non-existing) are chosen uniformly at random as training. For the baseline method AVER, we simply define the probability of a link using a modified version of Eq. \ref{equ:basicidp} without bias.
In Fig. \ref{fig:cold_start_classification} we compare our unified ML-IPM model (in solid red line) with (1) our model where weights are fixed (in regular dashed green line); (2) SL-IPM (in dotted blue line); (3) B-IPM (in dash-dotted black line); and (4) AVER (in bold dashed magenta line). We can see our unified approach is the best among all baseline methods, especially when the users are extremely sparsely linked.

\begin{figure*}[!t]
	\begin{center}
	\begin{minipage}{0.4\textwidth}
		\centering
		\includegraphics[width=\linewidth]{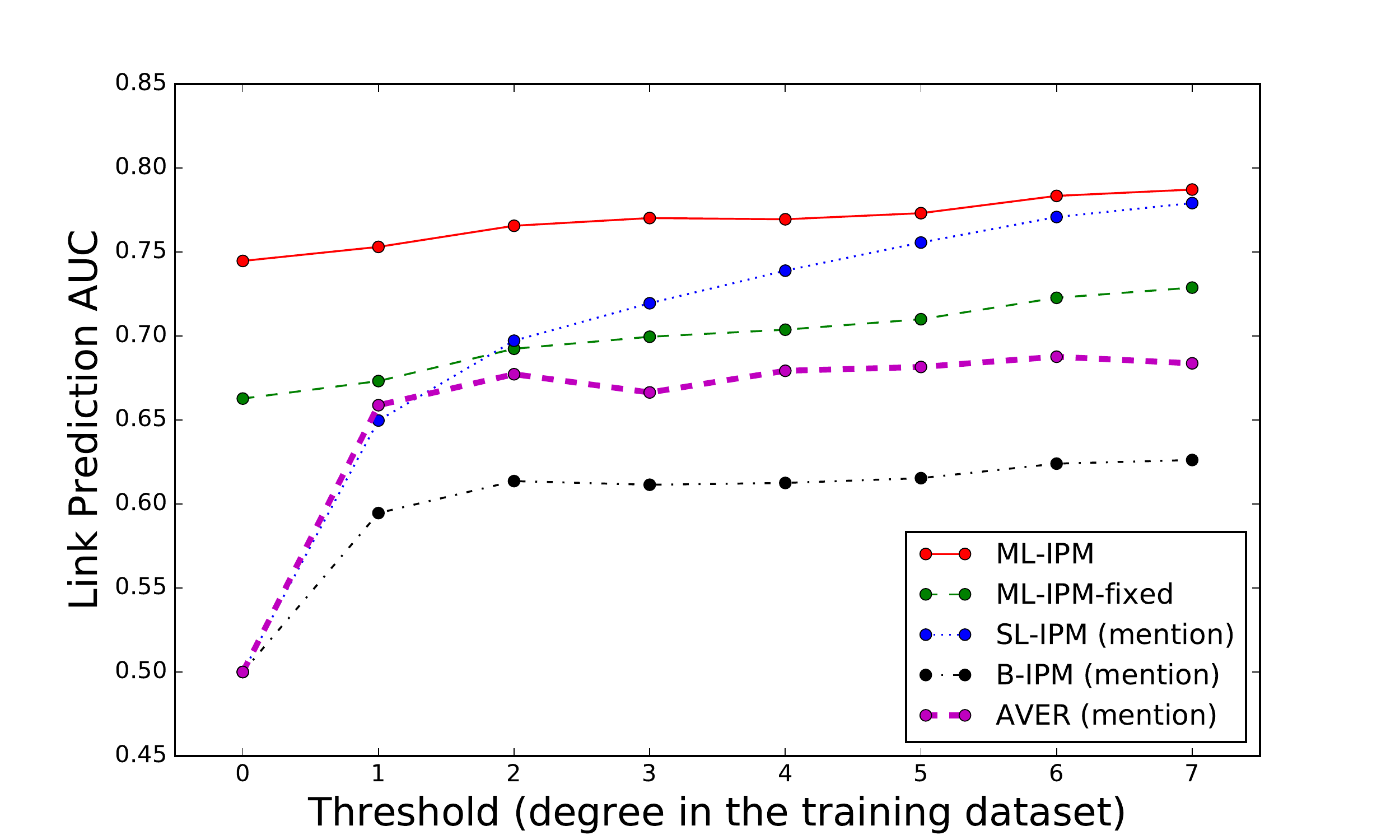} \\
		\textit{mention} network
	\end{minipage}%
	\begin{minipage}{0.4\textwidth}
		\centering
		\includegraphics[width=\linewidth]{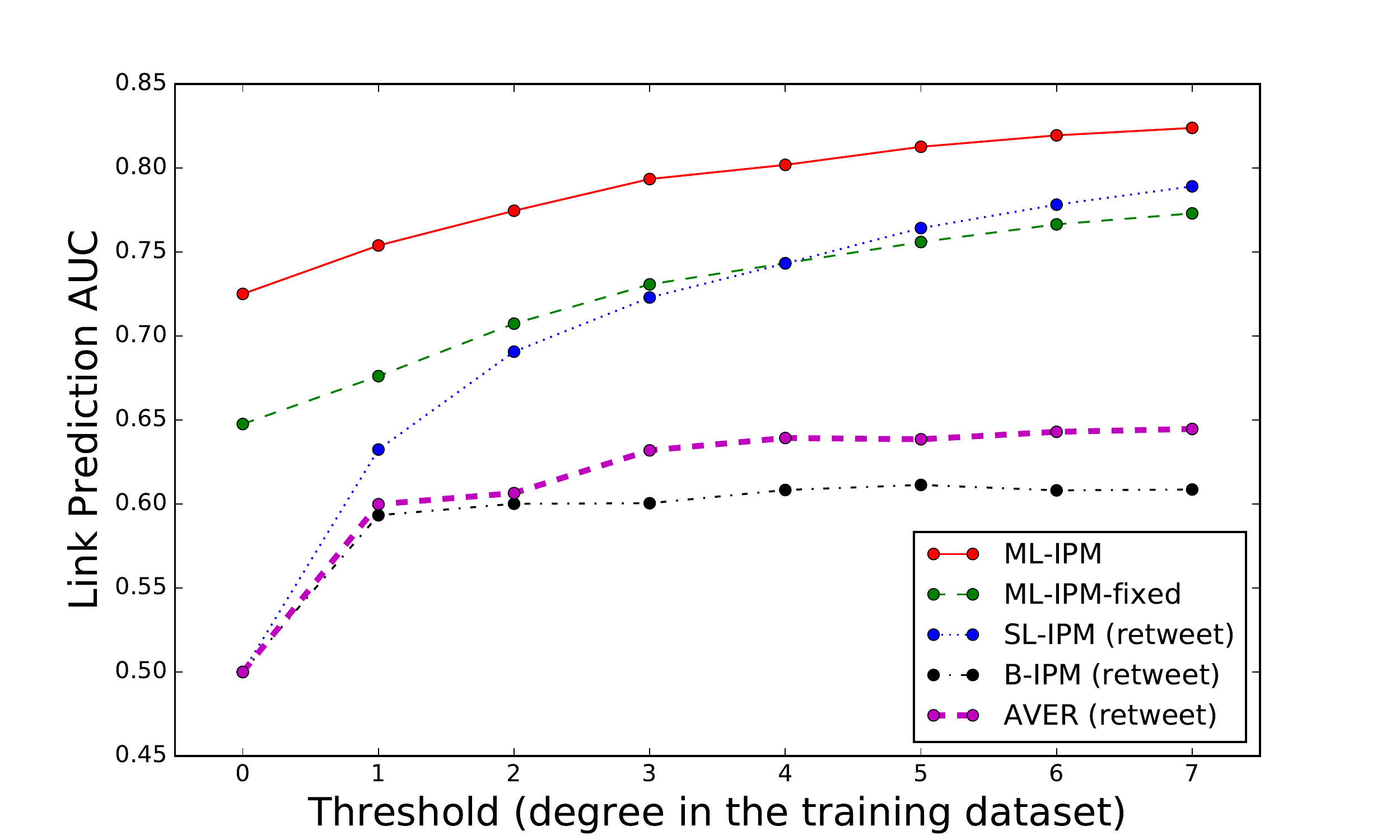}  \\
		\textit{retweet} network
	\end{minipage}%
	\end{center}
	\caption{Link prediction AUC for sparsely linked users in different types of networks.}
	\label{fig:cold_start_lp}
\end{figure*}

\subsubsection{Parameter Study}
Here we study how ideology dimension $K$ affects our model. The effect of $K$ is shown in Table \ref{table:idedim}. Our method is not sensitive to the choice of $K$, and the best $K$ should be around $5$. For above approaches and experiments, we use $K=5$ as the dimension of ideology, if not mentioned otherwise. 
\begin{table}[h]
	\centering
	\begin{tabular}{|c||c|c|c|c|} \hline
		Dimension $K$ & 1 & 3 & 5 & 7 \\ \hline
		Classification AUC & 0.979 & 0.981 & \textbf{0.986} & 0.977 \\ \hline
	\end{tabular}
	\caption{Effect of ideology dimension}
	\label{table:idedim}
\end{table}

\subsection{Case Studies}
We will analyze the relation weights $\bm{w} = \{w_r\}$ and visualize Twitter users' ideology in this subsection. 

It is interesting to study the relative weights $\bm{w}$ for different types of links. Intuitively, a larger $w_r$ implies the relation $r$ is more important in our task. The automatically learned weights are shown in Table \ref{table:netw}. We can observe the weight for \textit{retweet} is the highest among all types of links, while the weight for \textit{follow} is the lowest. This is reasonable as following someone on social network is common and does not stand for a very close relationship. For the other two relations, a retweet is likely to show the endorsement of the author of the original tweet; on the other hand, people are free to mention anybody and this behavior might indicate either pleasant or hostility. \delete{It's possible that a user encounters an unpleasant tweet and starts expressing different opinions by mentioning the author. }Therefore \textit{mention} behaviors may not indicate a strong homophily as \textit{retweet} does. In sum, a larger weight represents the significance of the corresponding source of information, and for the specific task of detecting Twitter users' ideology based on their social links, \textit{retweet} is the most crucial relation among all kinds of relations. 
\begin{table}[!htbp]
	\centering
	\begin{tabular}{|c||c|c|c|} \hline
		Relation $r$ & follow & mention & retweet \\ \hline
		Weight $w_r$ & 0.866 & 1.035 & 1.117 \\ \hline
	\end{tabular}
	\caption{Weights of different link types}
	\label{table:netw}
\end{table}

To further demonstrate the meaning of our learned weights, we randomly add 1,000,000 links between pairs of Twitter user and politician to our dataset. We denote this type of link as ``random'' and apply our model to the heterogeneous network with four relations (\textit{follow}, \textit{mention}, \textit{retweet} and \textit{random}). As a result, the weights of four types of links become 1.02, 1.23, 1.33, 0.59, respectively. The \textit{random} relation has the least importance score, and the trend of other three relations agrees with our previous result. This result shows our model is robust against noise and the learned weights reflect the importance of relations quite well.

\begin{figure}[H]
	\centering
	\subfigure[Ideology distribution for core users]{ \label{fig:core_users}
		\begin{minipage}[c]{0.45\textwidth}
		\centering
		\includegraphics[width=1.0\textwidth]{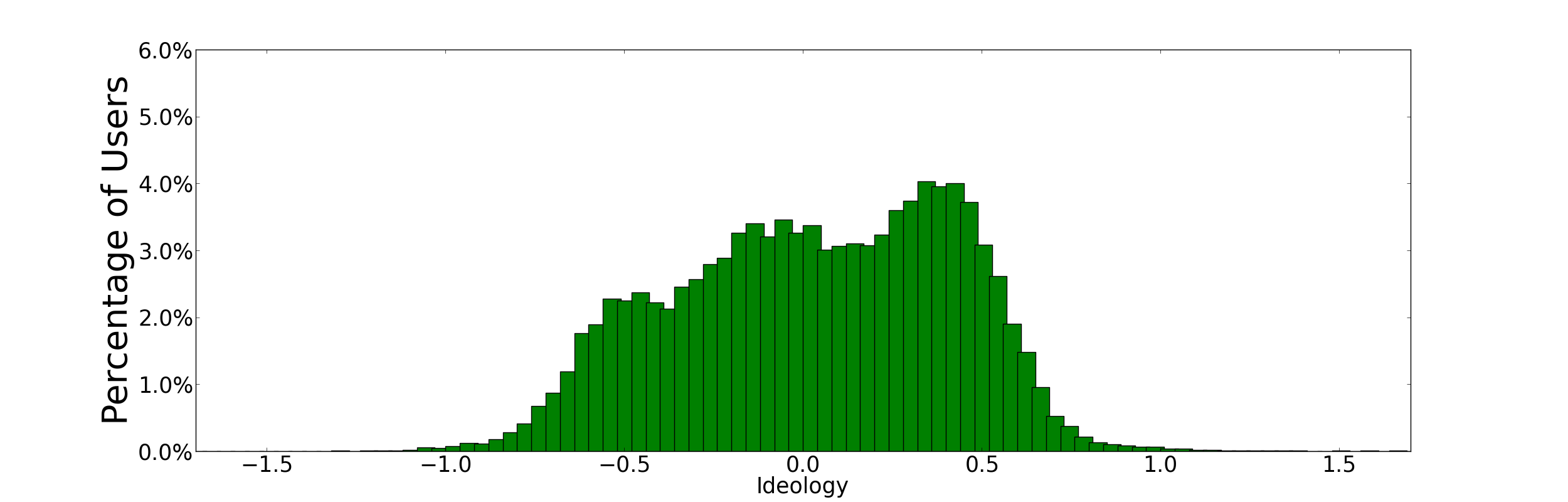}
		\end{minipage}%
	}
	\subfigure[Ideology distribution for peripheral users]{ \label{fig:pher_users}
		\begin{minipage}[c]{0.45\textwidth}
		\centering
		\includegraphics[width=1.0\textwidth]{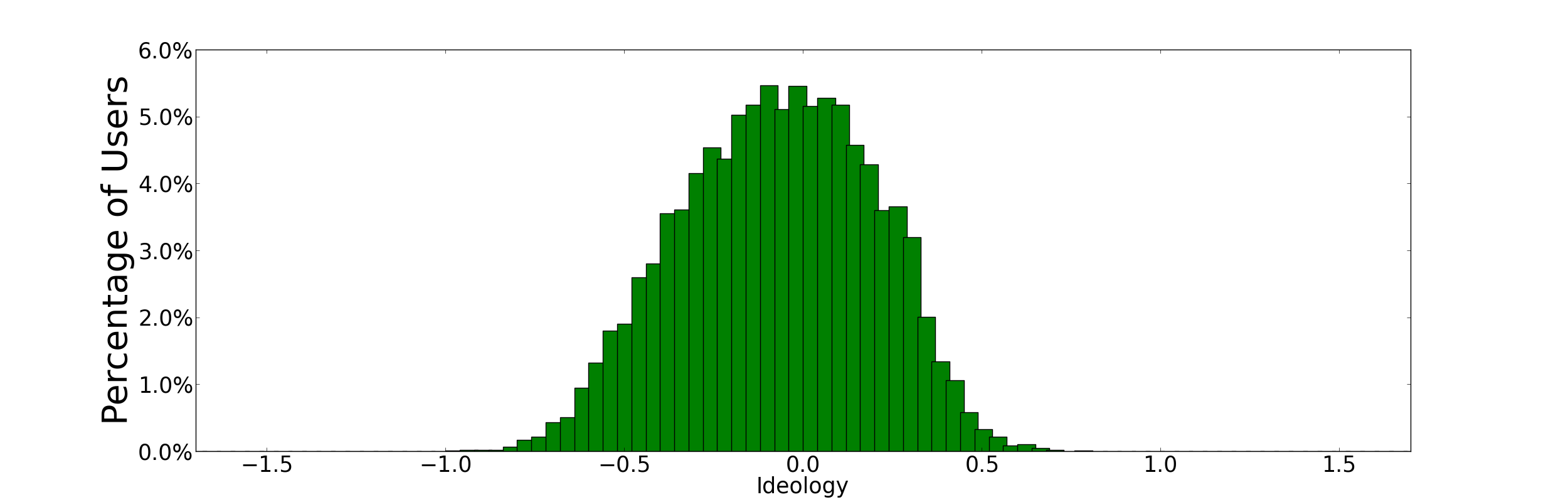}
		\end{minipage}
	}
	\caption{Ideology Distribution}\label{fig:all_users}
\end{figure}

In the remaining of this section we will visualize the latent ideology position of Twitter users. For visualization purposes we use $K=1$ in all figures below. Fig. \ref{fig:all_users} shows the distribution of ideology for all Twitter users in our dataset. We can observe three peaks around $x=-0.5$, $x=0$ and $x=0.5$ in Fig. \ref{fig:core_users}, which represent the three clusters of liberal, neutral and conservative users respectively. \delete{We may also interpret this plot as the mixture of three distributions, with the three peaks representing the mean of each. }Most of the users with less political interests (Fig. \ref{fig:pher_users}) will stay in the middle and form only one cluster. The statistics are reasonable because the distribution of peripheral users is more similar to the public, who are distributed more uniformly along the political spectrum. It is noteworthy that Twitter users are a biased sample of the public: they may be younger and more active on the Internet. Therefore skewness in our detected ideology may exist. 

\begin{figure}[!h]
	\centering
	\includegraphics[width=1.0\linewidth]{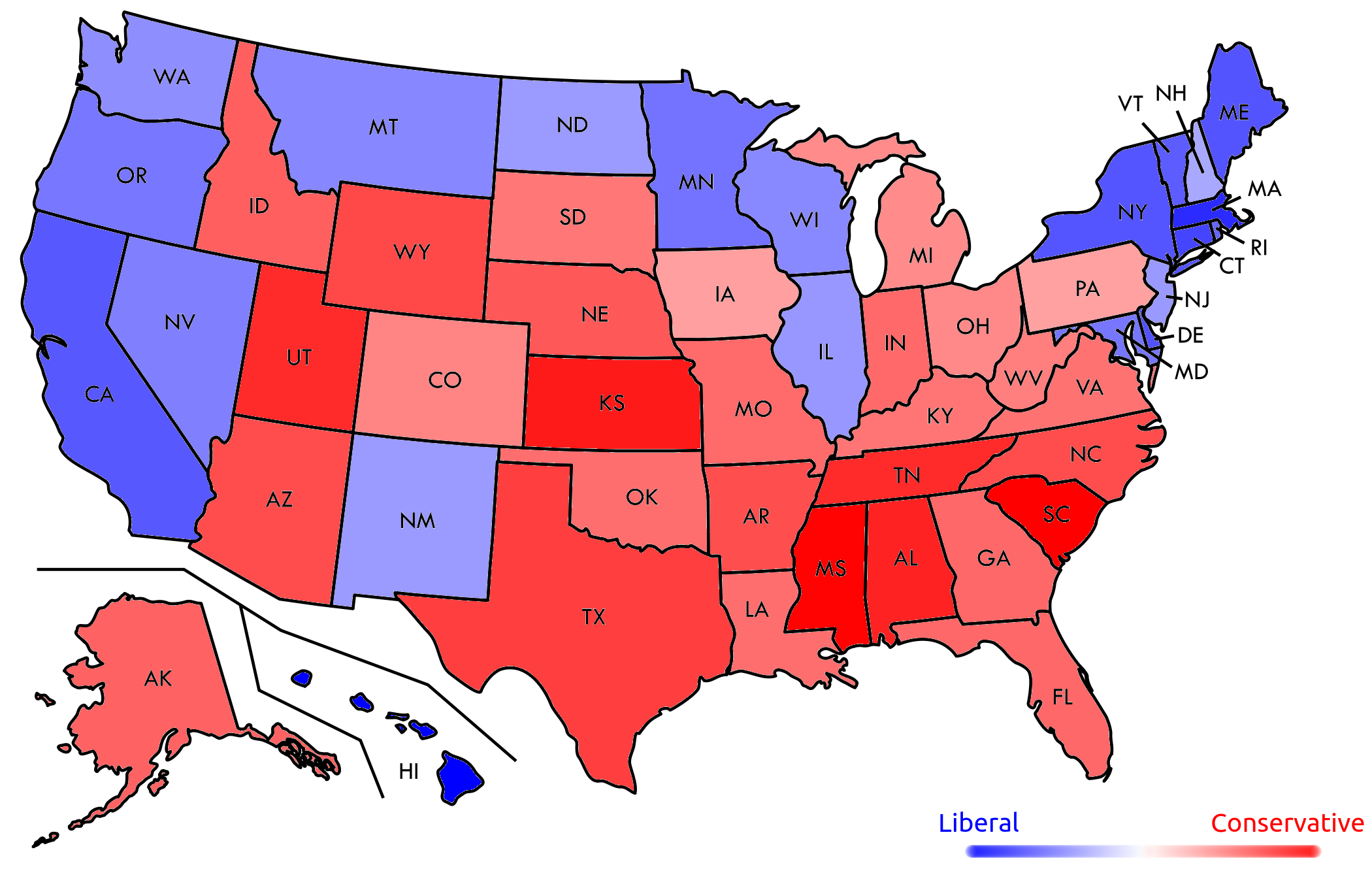}
	\caption{Average ideology for Twitter users in each state. Darker red means more conservative, while darker blue means more liberal.}
	\label{fig:state_map}
\end{figure}

Twitter users may also indicate their locations on their personal profiles. We collect all users in our dataset who claim themselves to live in one of the 50 states in the U.S. (or Washington, D.C.), and calculate the average ideology for each area. Then we are able to map the average score to a color between red and blue. As a result, 9,362 users are identified and 29 states are labeled as red (conservative), as shown in Fig. \ref{fig:state_map}. Note that our colored map may differ from political polls or election results, given the population of Twitter users and the sample variance in the dataset.

In Fig. \ref{fig:1dideal}, we also visualize the ideology for some popular media, magazine, newspaper accounts and a few political figures including president Barack Obama, Hillary Clinton and Donald Trump, the two presidential candidates in 2016\footnote{The ideology for Hillary Clinton and Donald Trump is estimated based on their most recent tweets and friends.} and so on. The length of each line (x-axis) represents the value of our detected ideology. The text on the y-axis denotes their Twitter screen name, and the color represents the political leaning provided by Mondo Times\footnote{\url{http://www.mondotimes.com/}, a worldwide news media directory.}: red represents a conservative account; blue stands for a liberal account; and black represents an account with no bias. From this vivid example we can see the position of those accounts on a unidimensional spectrum, and we are able to compare them with politicians. Again, there may be slight difference between the detected ideology and the actual political leaning, as the ideology is detected according to the behavior of these Twitter accounts instead of the editorial content. 

It is also worth analyzing the obvious exceptions in the results. Opposite to the label provided by Mondo Times, Wall Street Journal (WSJ) is considered as a media source leaning to the left in our study. When we investigate the social links of WSJ in our dataset, we find its links to politicians from two parties are balanced. Besides, our estimation refers to only the behaviors of their Twitter account (i.e. whom they follow/mention/retweet) instead of the editorial contents of the journal itself. This mainly explains the reason why our algorithm does not label it as a conservative account. Interestingly, WSJ is shown to be one of the most liberal online media outlets in some other studies \cite{groseclose2005measure,wong2013quantifying,lott2014newspaper}.

\nop{
\begin{table}[htbp]
	\centering
	\caption{Ideology for media accounts. C: conservative; L: liberal; N: no bias} \label{table:media}
	\begin{tabular}{|c|c|c|} \hline
		Media & \makecell{Detected \\ Ideology Score} 
		 & \makecell{Mondo Times \\ Classification} \\ \hhline{|=|=|=|}
		Newsmax Media & 0.612 & C \\ \hline
		The Washington Times & 0.300 & C \\ \hline
		Fox News & 0.260 & C \\ \hline
		U.S. News & 0.163 & C \\ \hline		
		Time & 0.148 & C \\ \hline
		CNBC & 0.042 & C \\ \hline
		CNN & 0.036 & N \\ \hline
		Forbes & 0.032 & C \\ \hline
		Foreign Policy & 0.006 & N \\ \hline
		USA TODAY & -0.024 & N \\ \hline
		Huffington Post & -0.036 & L \\ \hline
		POLITICO & -0.039 & N \\ \hline
		ABC & -0.042 & L \\ \hline
		National Public Radio & -0.088 & L \\ \hline
		Newsweek & -0.103 & N \\ \hline
		C-SPAN & -0.158 & N \\ \hline
		Wall Street Journal & -0.182 & C \\ \hline
		ProPublica & -0.301 & L \\ \hline
		CBS News & -0.303 & L \\ \hline
		MSNBC & -0.323 & L \\ \hline
		Washington Post & -0.326 & L \\ \hline
		Yahoo News & -0.331 & L \\ \hline
		Los Angeles Times & -0.392 & L \\ \hline
	\end{tabular}
\end{table}
}

\begin{figure}[!htbp]
	\centering
	\includegraphics[width=0.7\linewidth]{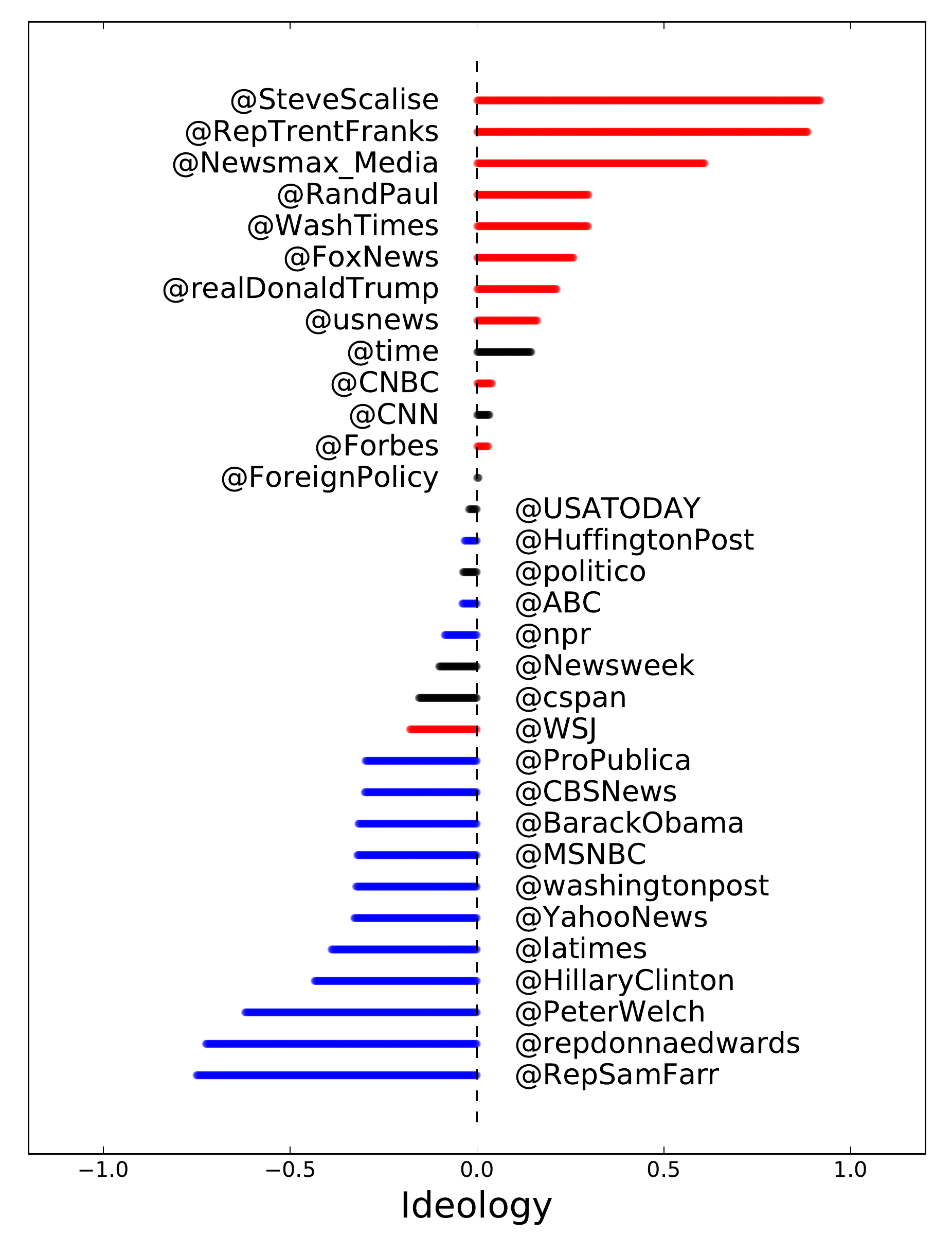}
	\caption{One dimensional ideology for selected users. Red: conservative; Blue: liberal; Black: no bias. Color labels are provided by Mondo Times.}
	\label{fig:1dideal}
\end{figure}

\nop{
\begin{figure}[htbp]
	\centering
	\includegraphics[width=0.92\linewidth]{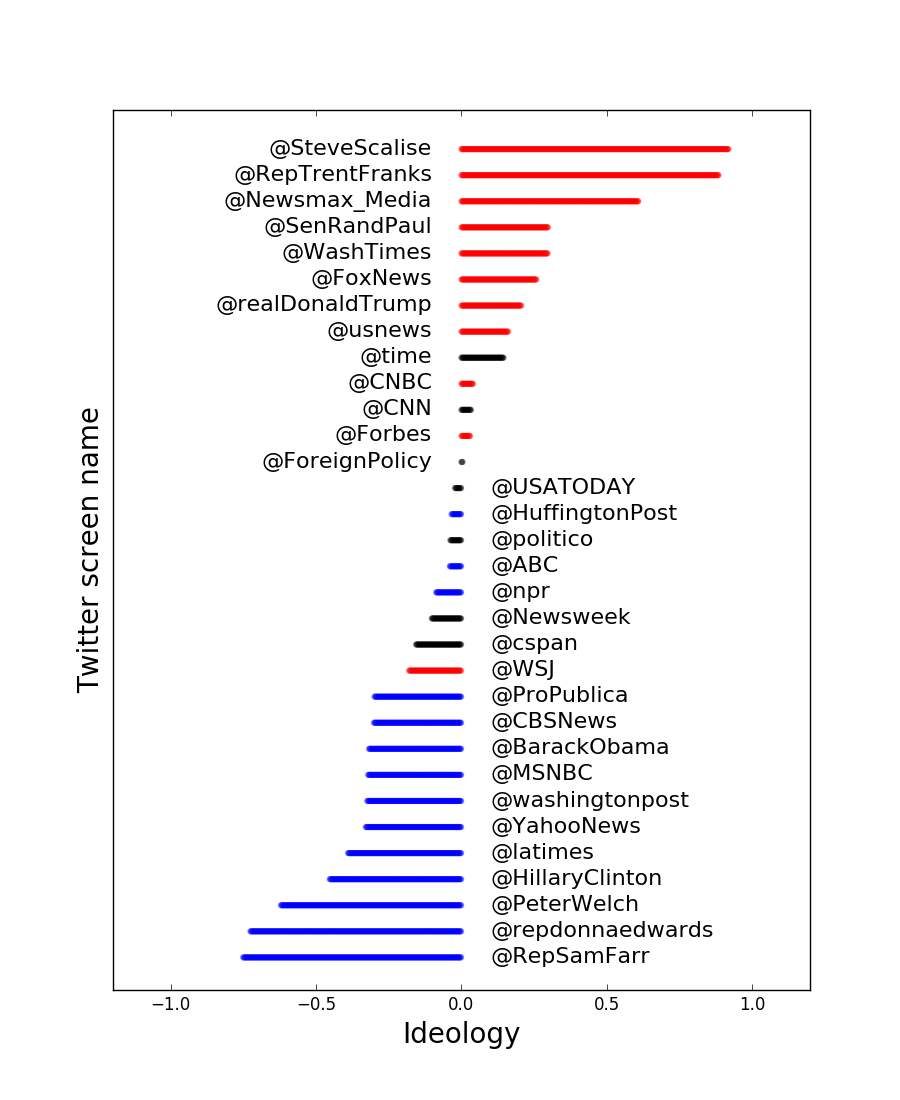}
	\caption{One dimensional ideology for selected users. Red: conservative; Blue: liberal; Black: no bias. Color labels are provided by Mondo Times.}
	\label{fig:1dideal}
\end{figure}
}

\section{Related Work} \label{sec:related}
\subsection{Ideology Detection in Roll Call Voting Data}
Ideal point models attempt to estimate the position of each lawmaker in the latent political space. Legislative voting is one of the sources for quantitative estimation of lawmakers' ideal points. Poole and Rosenthal \cite{poole1985spatial} were among the first few researchers in political science domain to provide a thorough and rigorous approach for ideology estimation. Afterwards, political scientists have proposed numerous methods to infer political ideology from roll call voting records \cite{poole1991,heckman1996linear,Poole1997,londregan1999estimating,jackman2001multidimensional,clinton2004statistical}. Researchers study the public voting record of lawmakers and model the probability of each vote, which is usually described as the interaction of the lawmaker's ideal point and the position of the bill. Along this line of research, computer science researchers extend the ideal point model to a variety of aspects, including applying natural language processing and topic modeling techniques on bills \cite{gerrish2011predicting,gerrish2012they,gu2014topic,iyyer2014political,nguyen2015tea}.

\subsection{Ideology Detection in Social Networks}
Apart from voting data, recently some approaches have been using information from social networks to analyze user's political leaning. These approaches can be further grouped into two categories according to the task they are dealing with: classification/clustering and continuous ideology estimation. Classification and clustering tasks predict which political party a user belongs to (or congregate users into several clusters), and can be handled from various aspects. When users interact with each other via messages, especially when they have discussions or arguments, methods from natural language processing can be applied to analyze the topics and people's sentiment. Pla and Hurtado \cite{pla2014political} analyze users' tweets on Twitter, extract features such as hashtags and punctuations, and determine the polarity of a tweet, and a user's political leaning is simply a weighted average of her tweets. \nop{There are also approaches which utilize information from the network structure and build node-level features as inputs for classifiers. }Boutet et al. \cite{boutet2012s} analyze the characteristics of three main parties in UK election, and predict which party a user supports by the amount of Twitter messages referring to political accounts. Based on debate records on various political topics, Gottipati et al. \cite{gottipati2013predicting} adopt a matrix factorization approach to discover users' attitudes towards different political issues, and use $k$-means on the user feature vector to congregate users into several clusters. Some researchers extract text features such as hashtags and latent semantic analysis of a conversion and feed them into a standard classifier in order to classify users \cite{rao2010classifying,conover2011predicting,pennacchiotti2011machine}. Users' public profile information (such as marriage status, age et al.) can also be utilized as inputs of classification models \cite{al2012homophily}. However, apart from classification tasks, it is also important to make inference about the continuous ideal points, which can be applied to rank ordering and demonstrate the relative difference between people in the same political party. Since a continuous measure of ideology is often desired for real world applications and our goal in the paper is to detect real-valued ideal points for users in social networks, those classification and clustering methods cannot be directly applied in our task. 

With respect to the work on estimating continuous political ideology, researchers have been mining the rich information in network structures to obtain an accurate ideology score for users on social networks. Typically, inference of a user's ideal point is made by exploring her neighbors and her relationship with labeled users (such as politicians whose party affiliation is clearly known to the public\nop{, or users who explicitly express their political leaning in their profiles}). Therefore, a simple yet intuitive approach would be calculating the ratio of the number of Democrats and Republicans that a user follows on the social network \cite{golbeck2011computing}. Wong et al. \cite{wong2013quantifying,wong2013media} assume liberal people tend to tweet more about liberal events and the same for conservative users. Given the political leaning of social events by labeling associated tweets, they are able to infer the ideology of users who have engaged. Barber{\'a} \cite{barbera2015birds} proposes a probabilistic model to describe the likelihood of the social network, in which the probability to observe a link between two users is defined as a function of their ideal points. \delete{Inference is then made using variational Bayesian methods.} Although most of the existing methods have utilized the homophily phenomenon \cite{mcpherson2001birds,wu2011says} in network generation and made appropriate assumptions of the network structure, they neglect the fact that most social networks\delete{are heterogeneous networks: namely more than one type of entities may exist and users can have multiple types of interactions.} contain multiple types of interactions between people. Those methods fail to define how close two nodes are when different link types are present. Although experts can pre-define weights for different link types, it is almost impossible to exhaustively list all possible combinations of link weights given the scale of real world data. Moreover, experiments show that the performance of existing approaches will suffer when a user is sparsely linked to others. Our approach, on the other hand, is able to automatically learn the weights for different types of links according to the network structure and determine how important each link type is in an ideology detection task. The cold start problem is also overcome by transferring information learned from one type of link to other link types.

\nop{
We summarize the related ideology estimation methods on social networks in Table \ref{table:related} for a quick reference. Our work lies in the lower right corner, and we are the only method that analyzes multiple types of links in heterogeneous networks. 
\begin{table}[h]
	\centering
	\tiny{
	\caption{Summary of ideology detection methods on social networks}\label{table:related}
	\begin{tabulary}{0.5\textwidth}{|c||c|c|} \hline
	\backslashbox{Data Source}{Task} & Classification/Clustering & Real-valued Ideology Estimation \\ \hline \hline
	Text or  User Profile
		& \makecell{Pla et al. \cite{pla2014political} \\ 
			Boutet et al. \cite{boutet2012s} \\
			Gottipati et al. \cite{gottipati2013predicting} \\ 
			Al Zamal et al. \cite{al2012homophily} \\
			Conover et al. \cite{conover2011predicting} \\
			Pennacchiotti et al. \cite{pennacchiotti2011machine} \\ 
			Rao et al. \cite{rao2010classifying} }
		&  N/A
		\\ \hline
	Links 
		&  \makecell{Wong et al. \cite{wong2013media} \\ 
			Wong et al. \cite{wong2013quantifying} }
		&  \makecell{Golbeck et al. \cite{golbeck2011computing} \\
			Barber{\'a} \cite{barbera2015birds} }
		\\ \hline
	\end{tabulary}
	}
\end{table}
}

\section{Conclusion and Future Work}

In this paper we present a novel approach for ideology detection on Twitter using heterogeneous types of links. Instead of predicting binary party affiliations to users, our work focuses on detecting the continuous ideal points for Twitter users, which is more comprehensive. In addition, we improve over traditional ideology estimation models by integrating information from heterogeneous link types in social networks. Specifically, our model is able to automatically update the importance scores of various relations on Twitter, and these weights are incorporated in the unified framework to achieve an outstanding performance. Moreover, our algorithm has an even better performance for sparsely linked users in separate networks. Finally, we evaluate our model on a subnetwork of Twitter, and the results show our advantage over the baseline methods. Extensive case studies also demonstrate our model's alignment with human intuitions.

One limitation of this work is that we only take links into consideration when deciding one's ideology. In the future, we plan to integrate other content information, such as text, into the current framework, to better understand people's ideology.

\bibliographystyle{abbrv}
\bibliography{ref}

\end{document}